\documentclass{epl}
\usepackage[dvips]{epsfig}
\usepackage{subfigure}
\usepackage{graphicx}
\def\be{\begin{equation}}
\def\ee#1{\label{#1}\end{equation}}
\title{ Accelerated expansion in bosonic and fermionic 2D
cosmologies with quantum effects}
\author{ L. L. Samojeden\footnote{samojed@fisica.ufpr.br},
G. M.  Kremer\footnote{kremer@fisica.ufpr.br}, and F. P. Devecchi\footnote
{devecchi@fisica.ufpr.br}}
\institute{Departamento de F\'\i sica, Universidade Federal do Paran\'a,
Caixa Postal 19044, 81531-990, Curitiba, Brazil}


\begin{document}

\maketitle

\begin{abstract}

In this work we analyze  the effects produced by bosonic and fermionic constituents, 
including quantum 
corrections, 
in two-dimensional (2D) cosmological models. We focus on a gravitational theory related to  the
Callan-Giddings-Harvey-Strominger model, to simulate the dynamics
 of a young, spatially-lineal, universe.
The cosmic substratum is formed by an {\it inflaton} field plus a matter component, sources
of the 2D gravitational
field;  the degrees of freedom also  include the presence of a dilaton field.
We show that this combination
permits, among other scenarios, the simulation of  a period of inflation, that would be 
followed by a
(bosonic/fermionic) matter dominated  era. We also analyse how  quantum effects  contribute to 
the
destiny of the expansion, given the fact that in 2D we have a consistent (renormalizable) 
quantum theory of 
gravity. The dynamical
 behavior
 of the system
follows from  the solution of the gravitational field equations, the  (Klein-Gordon and Dirac)
equations for the sources  and the dilaton field equation.
 Consistent (accelerated) regimes   are  present among the solutions of the 2D equations;
the results depend strongly  on the initial conditions
used for the dilaton field. In the particular case where  fermions  are included as 
matter fields a transition to a
decelerated  expansion is possible, something that does not  happen in the exclusively 
bosonic case.

\end{abstract}

\section {Introduction}

Cosmological models in lower dimensions have
been under analysis in several works ~\cite{RMann,Dese,K-D,CDKZ}.
These theories offer interesting mathematical results that,
if properly taken into account,  can also  be used    in realistic  models.
 These cosmological formulations are  obtained starting with different 2D
gravitational models\cite{ Bro, Dese,Cado}. As a first example, we can
take   the gravity
 model proposed by
  Teitelboim  and Jackiw~\cite{Jackiw,Teit,Bro} (JT model); a theory  that  provided
consistent results at   classical and quantum levels~\cite{Bro}.
This model   works as the ``closest counterpart''  of general
relativity in the 2D case; the obvious  candidate, Einsteinian gravity,
 furnishes no physics in 2D~\cite{CDKZ,RMann,Bro}.
As another alternative, one can consider the   CGHS model~\cite{CGHS};
here we have an additional  degree of freedom, the dilaton, giving the model the
 status of  2D analogue of  the Brans-Dicke theory \cite{Cado,CDKZ}. Here the cosmological results
include, as in the JT case, the
description of a matter/radiation filled Universe~\cite{RMann}. The
possibility of description of  inflation or
dark energy regimes depends  basically on  special
conditions (such as negative energy densities) or on the use  of the
van der Waals equation,    modeling the inflaton-matter  substratum\cite{CDKZ,RMann}. Another
investigations using these
models considered  the inclusion of quantum effects\cite{Koreans}; based on the remarkable fact that in 2D gravity is a
 renormalizable theory\cite{Bro}. The one-loop corrections
(integrating over scalar fields)
were calculated\cite{Bro,Koreans}, and the final  effective action encapsulates those effects
permitting semi-classical 2D cosmological regimes\cite{Koreans}. In an analogous manner it 
is possible
to consider fermions as sources; the   Dirac field equations are constructed using the   tetrad formalism
\cite{Marlo} in combination with the general covariance principle.  
In this work,
 we want  to
 analyze the  positive-accelerated
solutions of CGHS-inspired,  2D cosmologies with quantum corrections
 (without including a cosmological constant); 
these can be associated
to the description of an
 inflationary period followed by the beginning of a  (bosonic, fermionic) matter 
dominated period, with a compatible 
behavior of the physical quantities (scale factor and energy densities). We also
obtain an accelerated regime when  quantum
corrections
are considered. In fact, the behavior of the acceleration function depends
 strongly on the  initial conditions of the dilaton field, independently of the presence
of the quantum corrections. Another important point verified here is that it is possible to  obtain
naturally a transition to a decelerated (matter dominated) period when {\it fermions} are included as sources.

The manuscript is structured as follows: in Section II we make a brief
review of the dilatonic 2D model  with their cosmological applications. In Section III,
the  young 2D universe is described by taking into account the quantum corrections
of the field theory, when we have bosons as sources. In the last section fermions are included, playing the role
of matter constituents; this is  followed by our conclusions.   Units have been chosen so that
$G=c=h=1$.

\section{ II - Cosmology in  2D}

In this section we make a  short review  of 2D gravity
in a cosmological context focusing on the CGHS model mentioned in the introduction; for further details on 2D gravity 
and cosmology the authors refer to results presented
in \cite{RMann,Cado,CDKZ}.
2D gravity has the remarkable feature that when
the Einstein field
equations are invoked to rule the 2D space-time physics  no dynamics emerges~\cite{Bro}.
Several theories were proposed as alternatives, based on gauge principle grounds;
among them
the Jackiw-Teitelboim
(JT) model and the Callan-Giddings-Harvey-Strominger model (CGHS)\cite{Bro}. The CGHS model
was proposed initially
for the  investigation of
2D black-holes ~\cite{CGHS}.
The action in these models include  a  dilaton field, which plays a role that is
similar  to the one present in  the 4D Jordan-Brans-Dicke model\cite{Cado}.
The corresponding action was inspired in string theories~\cite{CGHS}, and
 the equations of motion that emerge are

\be 
G_{\mu\nu}=
e^{-2\phi}\left[R_{\mu \nu} -\Lambda g_{\mu\nu}- 2 \nabla _{\mu}
\nabla _{\nu} \phi \right] =
-8\pi T_{\mu \nu},\,\,\, 
R -\Lambda- 4(\nabla \phi )^2 + 4 \nabla ^2 \phi  = 0,
\ee{101}
where $\phi $ is the dilaton, $\Lambda$ is a cosmological constant and  $T_{\mu \nu}$ is the
energy-momentum tensor of the sources. The
Robertson-Walker metric has a simple  form in a 2D Riemannian space-time
$
ds^2=-(dt)^2+a(t)^2(dx)^2$,
where
$a(t)$  is the  cosmic scale factor, that  encloses
the  properties of the 2D gravitational field. In
fact, the usual geometrical quantities (Ricci tensor and curvature scalar) turn
out to be for this metric:
$
R_{00}={\ddot a\over a},\,
R_{11}=-\ddot a a,
\,
R=2{\ddot a\over a}.
$ The  regimes coming out include dust and radiation filled 2D Universes \cite{RMann,Cado}. 
The 
existence of positive accelerated
solutions depend, when barotropic equations of state are used, on the
imposition of  unusual
conditions like negative energy \cite {RMann, CDKZ}. However, interesting results appear 
when
the van der Waals (vdW) equation 
is considered \cite{CDKZ};
in fact a transition from  an inflationary  to a matter dominated 2D universe can be
obtained in this case, as far as the wdW equation approaches the behavior of a barotropic
expression as
a consequence of the accelerated expansion\cite{CDKZ}.
 On the other hand, the time
evolution of
  the cosmic  scale factor  and of its  acceleration can show, in some cases,  a dramatically
  different behavior when  compared to the one obtained in  the JT model (in its gauge fixed formulation)
 \cite{RMann,CDKZ}, as a consequence of 
the  dilaton field behavior in this model \cite{CDKZ}.  In all cases, after the inflaton-matter
 field transition occurs,  the accelerated regime  never returns\cite{CDKZ}. A fundamental point here
is that these results
arise from  classical formulations, and including only bosonic sources\cite{CDKZ, RMann}. It is interesting to
 consider quantum effects in these models,  given the fundamental point that in 2D gravity is
 renormalizable\cite{Bro}. 
In  the following sections we consider a  model inspired in the CGHS model, including bosonic 
and fermionic sources with quantum
effects, investigating the regimes  that emerge from   the 2D cosmology dynamics.

\section{ III - Bosonic 2D cosmology and Quantum effects}

Here we analyze the behaviour of a 2D cosmological theory where the gravitational sources are bosons, also 
investigating
 the influence of quantum effects in this CGHS based model (we don´t include the bulk cosmological
 constant present in the original theory\cite{CGHS, Koreans}).  The functional integration over one of the  sources
(the scalar field $f$, that will play the role of the inflaton)  furnishes
 the one-loop corrected action, that contains
 the usual conformal-anomaly term plus local contributions that are
included to maintain
the simple form of the
conserved currents \cite{Koreans}.
The corresponding total action is written as $S=S_{{\rm{cl}}} + S_{{\rm{qt}}}$,  with

\begin{eqnarray}
S_{{\rm{cl}}} = \frac{1}{2 \pi} \int\/d^{2}x\sqrt{-g}\left\{
	e^{-2\phi}\left[ R + 4(\nabla\phi)^2 \right]   -\frac12 \theta ^2 \chi ^2 f^2 - \frac12 (\nabla f)^2 - 
\frac12(\nabla \chi)^2 \right\},\label{s} \\
S_{{\rm{qt}}} = \frac{\kappa}{2\pi}\int d^2 x\sqrt{-g}\left\{
	-\frac{1}{4}R\frac{1}{\nabla ^2}R - \frac{\gamma}{2}\phi R + q (\nabla \phi )^2
	 \right\},
\end{eqnarray}

\noindent where $\phi$ is the dilaton and  $\kappa$  is the central charge related to the conformal anomaly 
\cite{Bro}; $\gamma$ and $q$ are constants that are usually chosen to obtain exact solvability \cite{Koreans}. 
In action ({4})
we have, besides the bosonic field $f$,  a matter constituent ( the scalar field $\chi$) whose mass will be 
proportional to the 
parameter $\theta $.

 After
localization of the quantum correction
(using an
auxiliary field
$\psi$ \cite{Bro, Koreans}), and plain use of the Hamilton principle,
 one obtains the gravitational equations of motion ($G_{\mu \nu}= T _{\mu \nu}$), which read explicitly
\begin{eqnarray}
 2 e^{-2\phi}\left[\dot{\phi}^2
	- \frac{\dot{a}}{a}\dot{\phi}  \right]= \frac14  \dot{f}^2 + \frac14  \dot{\chi}^2
	- \frac{\kappa}{2}\left(\frac{\dot{a}}{a}\right)^2
+ \frac{\gamma\kappa}{2}\frac{\dot{a}}{a}\dot{\phi}
-{\frac12} q\kappa\dot{\phi}^2 + \frac14 \theta ^2 \chi^2 f^2,\label{1}
\\
	 2 e^{-2\phi}\left[\ddot{\phi} - \dot{\phi}^2\right]=  \frac14\dot{f}^2 + \frac14 \dot{\chi}^2
	+ \kappa\left[\frac{\ddot{a}}{a}
	- \frac12\left(\frac{\dot{a}}{a}\right)^2\right]
	 - \frac{\gamma\kappa}{2}\ddot{\phi}
	-\frac12 q \kappa\dot{\phi}^2 - \frac14 \theta ^2 \chi^2 f^2.\label{2}
\end{eqnarray}

Analogously, we  obtain the  equations of motion for the dilaton $\phi$ and scalar fields ($f$ and $\chi$);
 these are 
  given by
\begin{eqnarray}
e^{-2\phi}\left(\frac{2\ddot{a}}{a} + 4\dot{\phi}^2 - 4\ddot{\phi}-
4\frac{\dot{a}}{a}\dot{\phi} \right)= -\frac{\gamma\kappa}{2}
\frac{\ddot{a}}{a}
        + q \kappa\left(\ddot{\phi} +
	\frac{\dot{a}}{a}\dot{\phi}\right),\\
\ddot f + \frac{\dot{a}}{a} \dot f+ \theta ^2 \chi ^2 f= 0 ,  \,\,\,\, 
 \ddot \chi  + \frac{\dot{a}}{a} \dot \chi+ \theta ^2 \chi  f^2= 0. \label{yy}
\end{eqnarray}
Here $f$ is   playing the role of the inflaton; as mentioned before $\chi$
is the matter component, and the term associated with $\theta$ represents the 
direct energy transfer 
between both fields.

Equations  (6), (7) and (8) are not linearly independent so we  use the following
combination of them
\begin{eqnarray}
\ddot \phi +{\dot a\over a}\dot\phi=2\dot\phi^2- {\frac 14} e^{2\phi} \theta ^2 \chi ^2 f^2 -
{\kappa\over4}e^{2\phi}\left[q\left(
 {\dot a\over a}\dot\phi+\ddot\phi\right)-\frac{\gamma}{2}{\ddot a \over a}\right],
 \\
 \dot \phi^2-{\ddot a\over 2a}=
 {\kappa\over4}e^{2\phi}\left[(\gamma-q)\left(
 {\dot a\over a}\dot\phi+\ddot\phi\right)+(\gamma-4){\ddot a \over 2a}\right]
 +{\frac 14} e^{2\phi} \theta ^2 \chi ^2 f^2.\label{yy1}
\end{eqnarray}
The classical  version of this model emerges  by simply considering $\kappa\equiv0$.
 As the expressions above show
the 2D universe dynamics  is ruled by  a highly non-linear, coupled system (6) -- (\ref{yy1}), having exact
solutions only in special
cases \cite{Koreans}. We proceed with a numerical integration of a
this system of differential
equations   in order to calculate  the energy densities of the inflaton $\rho_f$ and matter $ \rho_\chi$; 
we also obtain, accordingly, the evolution of the   cosmic scale factor $a$, of its acceleration,  and of the dilaton 
$\phi$. The inflaton 
and matter energy densities are defined by
\begin{eqnarray}
\rho _f =  \frac14 \dot {f^2}-{\kappa\over2}\left[\left({\dot a\over a}\right)^2
-\gamma\dot\phi{\dot a\over a}+q\dot\phi^2\right], \,\,\,\,
 \rho _{\chi} = \frac14 {\dot \chi}^2 +  \frac14 \theta ^2 \chi ^2 f^2  \label{G:yy2}\, .
\end{eqnarray}

\begin{figure}[h]
\center
\subfigure[ref1][ Scale factor   and ]{\includegraphics[width=5cm]{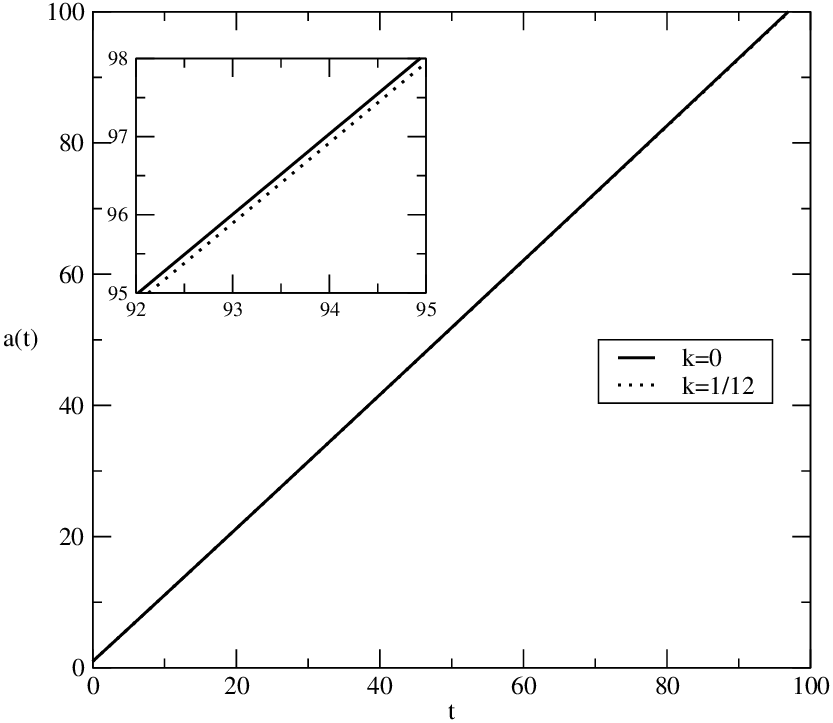}}
\qquad
\subfigure[ref2][acceleration vs time comparing  classical 
 with  quantum corrected solutions]{\includegraphics[width=5cm]{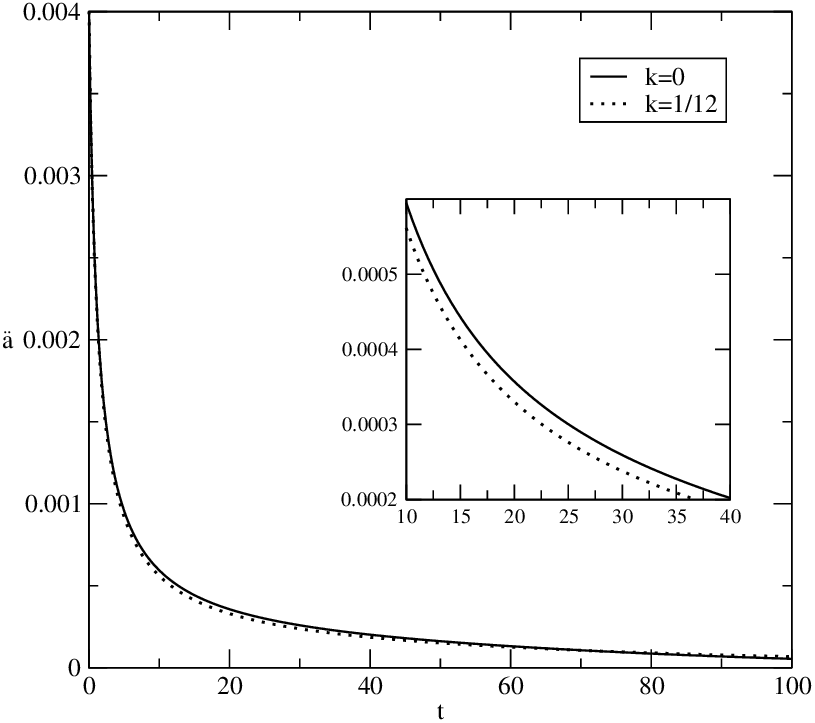}}

\end{figure}

The boundary conditions were chosen in such a manner that the universe is qualitatively seen as initially 
dominated by an 
inflaton field with initial energy density $\rho_f(0)=1$. The initial energy 
density of the matter field ($\chi$) 
was considered to be zero so that the matter is created during the evolution of the universe at expenses of the 
inflaton and gravitational fields.
The values adopted for the initial conditions were
$a(0)=1, \dot a(0)=1,  f(0)=0, \dot f(0)=2.03,$ ,  
$ \chi(0)=0.1, \dot\chi(0)=0, \phi(0)=0,  \dot\phi(0)=0.05.$ 
Moreover, the values chosen for the free parameters were: (a) quantum correction-terms $\kappa=1/12$, $q=2$ and
$\gamma = 6$ \cite{Koreans}; (b) coupling constant between the inflaton and matter fields
$\theta=10^{-3}$.

We start by describing the behavior of the scale factor $a(t)$. In a first
situation we consider the classical model, that
is, we take $\kappa=0$. In this case the time evolution of the scale factor depends
strongly on the initial value of
the time derivative of the dilaton field $\phi$ (as had  already been verified 
in \cite{CDKZ}). 
As those values increase the
expansion of the 2D universe becomes
faster. On the other hand, when we turn on the quantum effects they make the
expansion slower, although with the same
qualitative behavior present in the classical case (see figure a). Changes in
the initial value of the dilaton do not
produce sensible modifications in the scale factor evolution.

Next, we analyze  the behavior of the inflaton energy density $\rho _{f}$ (figure c)
and the matter field energy density
$\rho _{\chi}$ (figure d).
The first verification is that increasing values of $\dot
\phi (0)$ promote a faster decrease
of $\rho _{f}$ independently of the presence of the quantum terms in the
equations of motion; this is in tune
with the behavior of the the scale factor, confirming the expansion of the 2D
universe,  as far as we are working
with densities. For fixed values
of the initial conditions it is verified that the quantum effects furnish an
additional contribution to the energy
 density values, although without changing the qualitative character of the evolution
(see figure b).

\begin{figure}[h]
\center
\subfigure[ref1][ Inflaton energy density  and ]{\includegraphics[width=5cm]{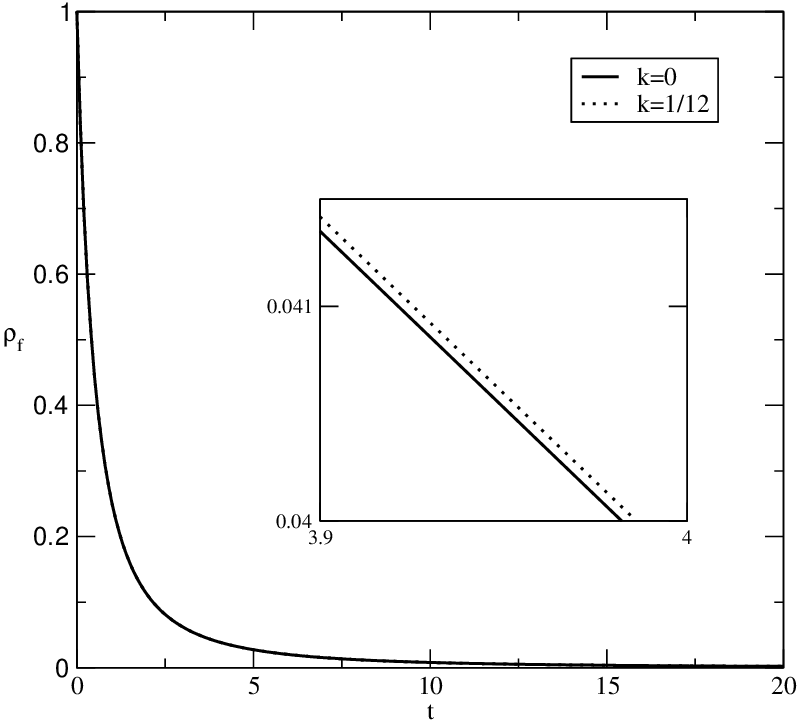}}
\qquad
\subfigure[ref2][matter energy density vs time comparing  classical 
 with  quantum corrected solutions]{\includegraphics[width=5cm]{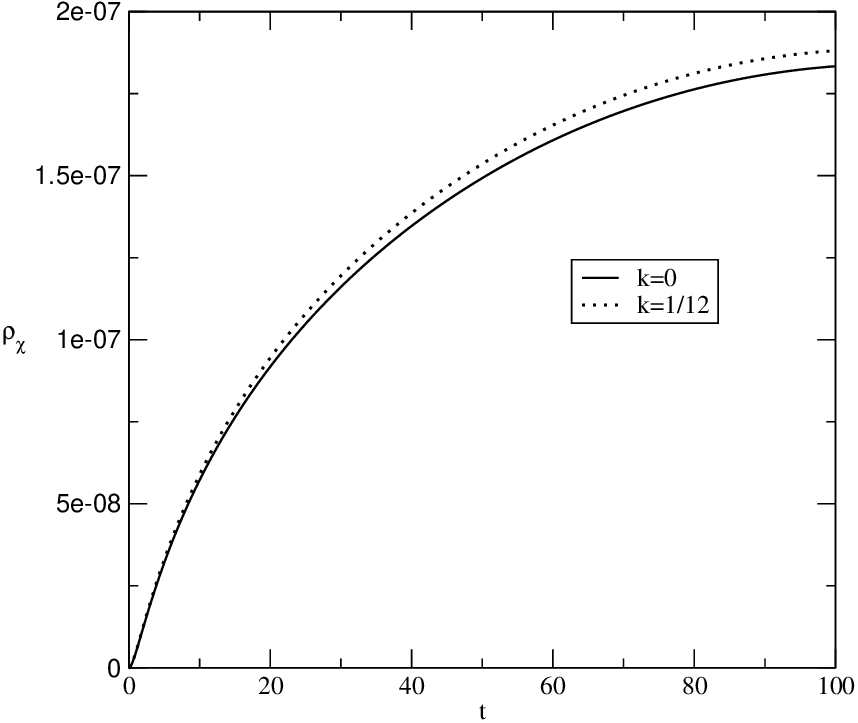}}

\end{figure}

 A similar effect
is confirmed for the  $\rho _{\chi}$ evolution, but the fundamental point
is that this matter density is increasing
with time;  the transfer of energy from the inflaton and
gravitational field contributions.  During its evolution
$\rho _{\chi}$ overcomes the inflaton density after a finite period of
time, but unlike what happens in the 4D
Einstein case this feature does not imply in a transition to a decelerated
regime as we discuss bellow (see fig b).

 The behavior of the acceleration $\ddot a$ depends strongly on
the initial condition for  derivative of the dilaton $\phi$, as was
 verified in \cite{CDKZ}.
In fact, in the classical case, increasing values of $\dot \phi (0)$  imply into a
smoother decay of $\ddot a$. When we
consider quantum effects there is a more drastic fall in the values of $\ddot a$, but, as a fundamental point,
only positive values are obtained, even when one is taking a long range of possibilities
for the initial conditions.
Besides, we also have that increasing values of $\dot \phi (0)$
promote a faster fall of the acceleration. As initial conclusions,  
we stress the fact that the inclusion of quantum effects
in this CGHS-inspired cosmology is responsible for an additional transfer of energy from the
 inflaton to the matter fields, promoting a faster transition to a matter dominated
2D universe, registered also by the fact  that $\ddot a$ assume lower values in that case. The model
where $q=0$ and $\gamma = 1$\cite{Koreans} was also considered but the results are not sensibly different from the 
$(q=2, \gamma=6)$
version presented above.

\section{ IV - 2D cosmology with  fermions }

In this section we  investigate the effects produced  by  a combination of  fermions 
 and bosons in a 2D cosmology, including quantum corrections. In this case the fermionic field represents a matter 
constituent and the bosonic field 
play, as in the previous model, the role of the inflaton.  The total action of the new theory is written  as
$
S_T =  S_{cl}  + S_{qt} + S_{F}, \label{}
$; where  the bosonic classical action  includes  now only one  source (the inflaton $f$). 
$S_{F}$ is the   (matter) fermionic action; it is given by

\begin{eqnarray}
S_{F} = \frac{1}{2\pi}\int d^2 x\sqrt{-g}\left\{\frac{i}{2}\left[\bar{\psi}\Gamma^{\mu}D_{\mu}\psi - 
(D_{\mu}\bar{\psi})\Gamma^{\mu}\psi\right]-V\right\}\label{}
\end{eqnarray}

\noindent  where in the last term  we consider an self-interacting  
fermionic potential $V(\psi , \bar \psi)$. 
The  gravitational equations of 
motion
$G_{\mu \nu}= T _{\mu \nu}$ include now 
  a fermionic term in the total energy-momentum tensor. From Noether theorem:
\begin{eqnarray}
T^F_{\mu\nu}=-\frac{i}{4}\left[\bar{\psi} \Gamma_{\mu}D_{\nu}\psi+\bar{\psi} \Gamma_{\nu}D_{\mu} 
\psi-D_{\nu}\bar{\psi}\Gamma_{\mu}\psi-D_{\mu}\bar{\psi}\Gamma_{\nu}\psi \right]
+g_{\mu\nu}\left\{\frac{i}{2}\left[\bar{\psi}\Gamma^{\alpha}D_{\alpha}\psi-D_{\alpha}\bar{\psi}\Gamma^{\alpha}
\psi \right]- V \right\}\label{}
\end{eqnarray}

\noindent The usual Dirac matrices ($\gamma^{\mu}$) become, due to the general covariance principle\cite{Marlo}:
\begin{equation}
\Gamma^0=\gamma^0,\,\, \Gamma^i=\frac{1}{a(t)}\gamma^i, \,\,
\Gamma^3=-\imath\sqrt{-g}\,\Gamma^0\Gamma^1\Gamma^2= \gamma^3.
\end{equation}
\noindent Following a  path analogous  to the one  in  section III, we obtain  two independent
  gravitational field equations, namely 

\begin{eqnarray}
\ddot \phi +{\dot a\over a}\dot\phi=2\dot\phi^2- \frac 14 e^{2\phi}\left[2V-\left(\bar{\psi}\frac{dV}{d\bar{\psi}}
+\frac{dV}{d\psi}\psi\right)\right]
-{\kappa\over4}e^{2\phi}\left[\gamma\left(
 {\dot a\over a}\dot\phi+\ddot\phi\right)-2{\ddot a \over a}\right], \label{}
 \\
 \dot \phi^2-{\ddot a\over 2a}=
 {\kappa\over4}e^{2\phi}\left[(\gamma-q)\left(
 {\dot a\over a}\dot\phi+\ddot\phi\right)+(\gamma-4){\ddot a \over 2a}\right]
+\frac 14 e^{2\phi}\left[2V-\left(\bar{\psi}\frac{dV}{d\bar{\psi}}+\frac{dV}{d\psi}\psi\right)\right].\label{}
\end{eqnarray}

\noindent The dilaton and bosonic fields  evolution is ruled   again by  expressions (8) 
and (9), with $\theta =0$.  For the 
fermionic field we have the curved space-time  Dirac equations\cite{Marlo} 
$
\dot{\psi}+\frac12 {\dot a\over a}\psi + i\gamma^0\frac{dV}{d\bar{\psi}}=0,
$ with an analogous expression for $\bar\psi$.
\noindent Using the explicit form of the fermionic potential $V=(\bar{\psi}\gamma^3\psi)^{2n}$ and the 
bi-spinor
 components ($\psi_A, \psi_B$), we can write the field equations  more coveniently as

\begin{eqnarray}
\ddot \phi +{\dot a\over a}\dot\phi=2\dot\phi^2-{\kappa\over4}e^{2\phi}\left[\gamma\left(
 {\dot a\over a}\dot\phi+\ddot\phi\right)-2{\ddot a \over a}\right] 
+ \frac{1}{2} e^{2\phi}(2n-1)(i\psi^{\ast}_{B}\psi_{A}-i\psi^{\ast}_{A}\psi_{B})^{2n}, \label{}
 \\
 \dot \phi^2-{\ddot a\over 2a}=
 {\kappa\over4}e^{2\phi}\left[(\gamma-q)\left(
 {\dot a\over a}\dot\phi+\ddot\phi\right)+(\gamma-4){\ddot a \over 2a}\right]
-\frac 12 e^{2\phi}(2n-1)(i\psi^{\ast}_{B}\psi_{A}-i\psi^{\ast}_{A}\psi_{B})^{2n},\label{}
\\
\dot{\psi_{A}}+\frac 12 {\dot a\over a}\psi_{A} + 2n(i\psi^{\ast}_{B}\psi_{A}-i\psi^{\ast}_{A}\psi_{B})^{2n-1}\psi_{B}=0, 
\end{eqnarray} with an analogous expression for $\psi_B$.
The initial  conditions were chosen, again,  in such a manner that the 2D universe starts dominated by 
an inflaton field with initial energy density $\rho_{f}(0)=1$. The initial energy density of the fermionic 
 field was fixed   so that  matter is created at expenses of the inflaton and gravitational fields 
energy (we do not consider a direct interaction between the sources in this case).
The values adopted for these initial conditions were
$ {a(0)=1,\quad\dot a(0)=1,\quad \phi(0)=0,\quad \dot \phi(0)=-0.01,
 f(0)=0, \quad \dot f(0)=2.0,\quad \psi_A(0)=-0.1i, \psi_B(0)=0.01} .
$

\noindent  The values chosen for the quantum correction  parameters coincide with those of the previous section  
 $\kappa=1/12$, $q=2$ and
$\gamma = 6$ \cite{Koreans}. We start the numerical analysis by describing the behavior of the 2D universe when the 
power ($n$) in the fermionic potential $V(\bar \psi, \psi)$ was chosen to be $n=1/2$ and $n=1$. These values 
correspond to special cases of the Nambu-Jona-Lasinio potential \cite{Nambu}. What is verified in these
 cases  is a behaviour completely analogous to the  strictly bosonic model, showing no distinct features in the 
classical
 nor in the quantum corrected versions. An important point here is that there is no transition in the
 acceleration evolution, meaning that an entrance in a matter dominated/decelerated period (following inflation) is
 not possible in this case.

The most interesting result appears, however,  when we choose the power $n$ to be in the neigborhood  of values
 $n\approx 0.49$. 
Starting with  the behavior of the scale factor $a(t)$, making initially $\kappa=0$,  what we verify is that 
we have an 
 analogous situation to  the strictly bosonic case (see figure 1); we have a fast expansion and  this  depends
strongly on the initial value of
the time derivative of the dilaton field $\phi$. When we turn on the quantum effects 
 the expansion becomes slower, again,  as in the previous model.

Next, we analyze  the behavior of the fermion energy density $\rho _{\psi}$ 
and the inflaton energy density 
$\rho _{f}$
that are given by 
$
\rho _{f} =  \frac14 \dot {\chi^2}-{\kappa\over2}\left[\left({\dot a\over a}\right)^2
-\gamma\dot\phi{\dot a\over a}+q\dot\phi^2\right], \,\,\,\,
\rho _{\psi} = (i(\psi^{\ast}_{B}\psi_{A}-\psi^{\ast}_{A}\psi_{B}))^{2n}. 
$
The inflaton density behaves again as in figure c; the fundamental point here
is that the  fermionic matter   density is  evolving due to the energy transference from the
 inflaton (via gravitational field). On the other hand, it is decreasing in absolute values
 due to the quick expansion of the 2D universe (figure 1).  The quantum effects promote an additional 
transference
 of energy from the inflaton to the  matter field, like in the strictly bosonic case.

The behavior of the acceleration  follows the bosonic case patterns when one focus on the dilaton field: 
increasing values of $\dot \phi (0)$  imply into a
smoother decay of the acceleration. When we
consider quantum effects there is a more drastic fall in the values of $\ddot a$. The most important feature of 
the case 
$n=0.49$ is that we have a transition from the accelerated regime to a decelerated period (see figure e) showing 
that the 
inclusion of fermionic matter as a constituent permit a gradual  exit form the inflation period. Besides, 
the quantum contributions  decay as consequence of the  2D universe expansion, approaching naturally  strictly 
classical dynamics.

\begin{figure}[h]
\center
\subfigure[ref2][acceleration vs time comparing  classical 
 with  quantum corrected solutions]{\includegraphics[width=5cm]{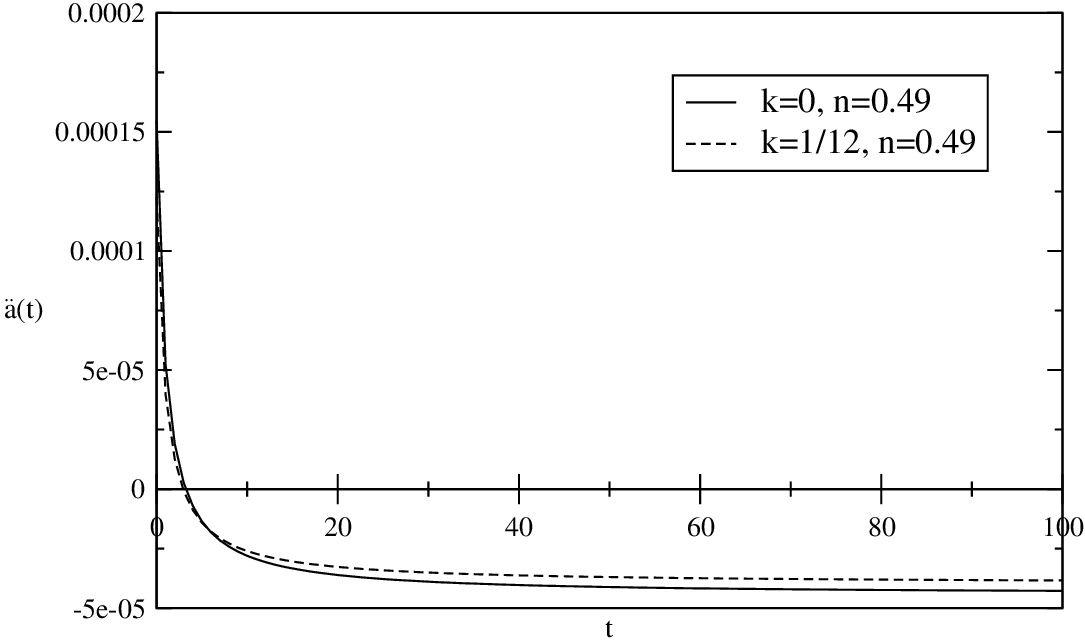}}
\end{figure}

Our final comments
 stress the fact that the inclusion of quantum effects
in this 2D cosmology is responsible for an additional transfer of energy from the
 inflaton to the matter  fields, promoting a faster transition to a matter dominated
2D universe, although always with positive acceleration in the bosonic case.   
In the fermionic case we verify that the inclusion of the curved space-time Dirac dynamics  
is responsible  for an interesting 
transition to a
 {\it decelerated} regime, dominated by fermionic matter.   Again,the model
where $q=0$ and $\gamma = 1$ was also considered but the results are not sensibly different 
from the $(q=2,\gamma=6)$
case.

\acknowledgments
 FPD and GMK acknowledge the support by
 CNPq-Brazil.

\end{document}